\documentclass[aps,prl,twocolumn,superscriptaddress,groupedaddress]{revtex4}%for review and submission
\usepackage{graphicx,amsmath,amssymb,amsfonts,latexsym,color,dcolumn,bm,mathtools}

\newcommand{\beq}{\begin{equation}}
\newcommand{\eeq}{\end{equation}}
\newcommand{\bea}{\begin{eqnarray}}
\newcommand{\eea}{\end{eqnarray}}
\providecommand{\abs}[1]{\left\lvert#1\right\rvert}

\providecommand{\bra}[1]{\langle #1 \rvert}
\providecommand{\ket}[1]{\lvert #1 \rangle}
\providecommand{\bbra}[1]{\langle #1 \rvert\rvert}
\providecommand{\kket}[1]{\lvert\lvert #1 \rangle}

\providecommand{\so}[1]{#1}

\newcommand{\ud}{\mathrm{d}}

\graphicspath{ {/} }

\usepackage{tikz}
\usepgflibrary{arrows}
\usetikzlibrary{decorations}
\usetikzlibrary{decorations.pathmorphing,patterns}
\usetikzlibrary{scopes}
\usetikzlibrary{shapes, matrix}
\begin{document}

\title{Non-linear Fano Interferences in Open Quantum Systems: an Exactly Solvable Model} 
\author{Daniel Finkelstein-Shapiro}
\affiliation{Department of Chemistry and Biochemistry, Arizona State University, Tempe AZ 85282}
\affiliation{Sorbonne Universit\'es, UPMC Univ Paris 06, UMR 7616, Laboratoire de Chimie Th\'eorique, F-75005, Paris, France}
\affiliation{CNRS, UMR 7616, Laboratoire de Chimie Th\'eorique, F-75005, Paris, France}
\author{Monica Calatayud}
\affiliation{Sorbonne Universit\'es, UPMC Univ Paris 06, UMR 7616, Laboratoire de Chimie Th\'eorique, F-75005, Paris, France}
\affiliation{CNRS, UMR 7616, Laboratoire de Chimie Th\'eorique, F-75005, Paris, France}
\affiliation{Institut Universitaire de France, France
}
\author{Osman Atabek}
\affiliation{Institut des Sciences Mol\'eculaires d'Orsay, B\^atiment 350, UMR8214, CNRS-Univ. Paris-Sud, Univ. Paris-Saclay 91405 Orsay, France}
\author{Vladimiro Mujica}
\affiliation{Department of Chemistry and Biochemistry, Arizona State University, Tempe AZ 85282}
\author{Arne Keller}
\affiliation{Institut des Sciences Mol\'eculaires d'Orsay, B\^atiment 350, UMR8214, CNRS-Univ. Paris-Sud, Univ. Paris-Saclay 91405 Orsay, France}

\begin{abstract}
We obtain an explicit solution for the stationary state  populations  of a dissipative Fano model, where a discrete excited state is coupled to a continumm set of states; both excited set of states are reachable by photo-excitation from the ground state. The dissipative dynamic is described by a Liouville equation in Lindblad form and the field intensity can take arbitrary values within the model. We show that the continuum states population as a function of laser frequency can always be expressed as a Fano profile plus a Lorentzian function with effective  parameters whose explicit expressions are given in the case of a closed system coupled to a bath as well as for the original Fano scattering framework.
Although the solution is intricate, it can be elegantly expressed as a linear transformation of the kernel of a $4\times 4$ matrix which has the meaning of an effective Liouvillian. We unveil key notable processes related to the  optical nonlinearity and  which had not been reported to date: electromagnetic induced-transparency, population inversions, power narrowing and broadening, as well as an effective reduction of the Fano asymmetry parameter. 
\end{abstract}

\maketitle

\section{Introduction}
The original Fano model was introduced by U.~Fano in
1935~\cite{Fano1935} and formalized in 1961 \cite{Fano1961} to explain the asymmetry in the  absorption or photo-current profile as a function of laser frequency used to ionize a gaz of Helium-like atoms \cite{Beutler1935}. Previously, similar diffuse absorption bands induced  by iodine-chloride pre-dissociation have been observed and theoretically addressed \cite{Rice1933}. Friederichs~\cite{Friedrichs1948} developed the mathematical formalism of perturbation of linear operators to describe the essential feature of the Fano model: a discrete state coupled to a continuum set of states; both sets of
states being reachable by photo-excitation from the ground state. The resulting photo-current, which is proportional to the population of the continuum set of states, as a function of the excitation laser
frequency $\omega_L$ is known as the Beutler-Fano or Fano profile:
\begin{equation}
f(\epsilon;q)=\frac{(q+\epsilon)^2}{\epsilon^2+1},
\label{eq:Fano-classic}
\end{equation}
where $q$ is the ratio of the transition dipole moment of the ground-discrete and ground-continuum transitions, and $\epsilon=(\omega_{L}-\omega_e)/\gamma$ where $\hbar\omega_e$ is the energy of the discrete state relative to the ground state, $\omega_L$ is the incident radiation field frequency, and $\hbar\gamma=n\pi V^2$ is the linewidth of the excited state, induced by its coupling (per unit of energy) $nV^2$ to the continuum set of states, $n$ being the density of states.
The Fano literature is extensive and we cannot do justice in this paper to all the contributions since 1935. The interested reader is pointed to Refs. \cite{Fano2012,Satpathy2012,Rau2004}. 

Two important extensions of the original model have been considered: inclusion
of incoherent relaxation and dephasing processes and high field intensities. The motivation to include incoherent processes was first to describe the pressure broadening~\cite{Fano1963} due to elastic collisions,  laser phase fluctuations~\cite{Eberly1983} and spontaneous emission~\cite{Agarwal1982,Robicheaux1995}. 
Nowadays, Fano profiles in  nanoscale structures are standard \cite{Mirosh2010,Lucky2012}, for example in plasmonic nanostructures \cite{Pakizeh2009,Lucky2010,Hsu2014}, quantum dots, decorated nanoparticles~\cite{Lombardi2010} and spin filters \cite{Song2003}. The coherent coupling with the incident light  induces large Rabi frequencies which in turn compete with the relaxation rates in order to modify the stationary state.  The ability to predict the lineshape, and in general  to investigate non linear optical phenomena in the presence of a continuum set of states are the main motivations to consider  large incident field intensity. 
A growing number of experimental and theoretical papers have been appearing on the subject of coherent control and ultrafast pulses on Fano models \cite{Ott2013,Chu2010,Wickenhauser2005,Loh2013}. 

In spite of the ubiquity of the Fano interferences, to the best of our knowledge, explicit solutions for high laser
intensities and including general relaxation processes have not been
obtained. Even in Refs.~\cite{Kroner2008, Zhang2011}, dealing with
quantum dots, despite some approximations no analytical expressions
are derived that afford a simple physical interpretation of the results.

In a recent work~\cite{PRL1}, we investigated the signatures of the Fano interferences in the emitted spectrum of a system with a vibrational manifold.
We obtained explicit expressions of  spectroscopic
observables like Rayleigh, Raman and fluorescence emission but
restricted to the low intensity field limit where the lowest order of perturbation was enough to describe the laser-matter interaction.
In this letter, we focus on the description of the non-linear Fano effect on the total population of the continuum excited state, the observable in the original Fano model,  but for the general case of large intensities of the laser field and including dissipation processes. 
Here, unlike Ref.~\cite{PRL1}, obtaining explicit expressions requires non perturbative calculations. 
We present a method that allows us to obtain such an 
explicit formulation for large field strengths and Markovian baths
in an elegant and intuitive framework where the entire solution is
formulated in terms of a $4\times 4$ matrix corresponding to an
effective Liouvillian acting in the space of the discrete states only. 
 We calculate  the usual Fano observable,
 that is, the total population of the continuum set of states,  which is related to the optical
 absorption or to the  photo-current, as a
 function of the incident laser frequency. The striking result is that
 such a function can be written exactly as  a linear
 combination of a Fano
profile and a Lorentzian function like in Ref.~\cite{PRL1}, but where
the Fano $q$ and $\epsilon$ coefficients become effective
parameters that are
functions of  the field intensity
and the decay rates.

\section{Model}
\begin{figure}[ht]
\centering
\begin{tikzpicture}[scale=0.7]
%ground state
\draw[thick] (0,0) -- (4cm,0) node[right]{$\ket{g}$};
%excited state
\draw[thick] (0,3cm)--(4cm,3cm) node[right]{$\ket{e}$};
%continuum
\draw[fill=gray]  (6cm,1cm) rectangle (7cm,5cm) node[right]{$\ket{k}$} ;
% excitation discrete
\draw[->,thick] (2cm,0) --(2cm,2.9cm) node[midway, right] {$\mu_{e}$};
% excitation continuum
\draw [->,thick] (2.5cm,0cm)--(6.25cm,3.25cm)
node[midway, above] {$\mu_{k}$};
%Vcoupling
\draw[<->,thick] (3cm,3.1cm) to[out=45,in=135] node [sloped, above] {$V(k)$} (6.5cm,3.1cm);
%\relaxation
\draw [->,thick,decorate,decoration=snake] (6.25cm,2.75cm)--(3cm,0)
node[midway,sloped ,below,] {$\Gamma(k)$};
\draw [->,thick,decorate,decoration=snake] (1cm,2.9cm)--(1cm,0) node[midway,left] {$\Gamma_e$};
\end{tikzpicture}
\caption{\label{fig:FanoDissip} Energy levels and transitions of a Fano-type model with dissipation (see main text for notations)}
\end{figure}
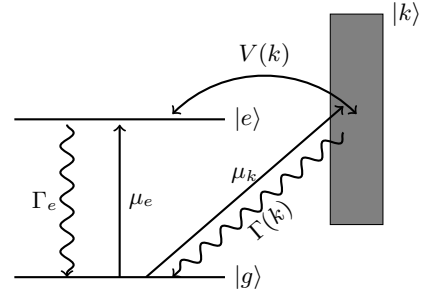

%\textbf{Introduction of the Hamiltonian and Dissipation
%Superoperator}. 
Although the objectives and results of the present work differ from our
previous one~\cite{PRL1}, the model  is similar. As there are
some differences and for the sake of introducing the notation in a self-contained way, we summarize it below.
The ingredients of the model are schematically presented in Fig.~\ref{fig:FanoDissip}.
The Hamiltonian $H=H_0+H_V+H_F $ is exactly the same as in the original Fano
model~\cite{Fano1961}:
\begin{align}
%\begin{split}
&H_0=E_{0}\ket{g}\bra{g}+E_e\ket{e}\bra{e}+\int dk \epsilon_k\ket{k}\bra{k} \nonumber \\
&H_V=\int dk V(k)\ket{e}\bra{k}+V^*(k)\ket{k}\bra{e} \nonumber \\
&H_F=F \left[\mu_{e}\cos(\omega_L t)\ket{g}\bra{e}+\mu_{e}^*\cos(\omega_L t)\ket{e}\bra{g}\right] \nonumber \\
&+F \int dk \left[\mu_{k}\cos(\omega_L t)\ket{g}\bra{k}+\mu_{k}^*\cos(\omega_L t)\ket{k}\bra{g}\right],
%\end{split}
\label{eq:Hamiltonian}
\end{align}
where $H_0$ is the site Hamiltonian, $H_V$ is the coupling of the excited state to the continuum. For simplicity, in the following,  we will consider that $V(k) = \bra{e}H_V\ket{k}$ is real. $H_F$ is the interaction with the incident radiation field, allowing transitions from the ground state to the discrete excited state $g \leftrightarrow e$ and to the continuum of states $g \leftrightarrow k$, $\mu_{ij}=\bra{i}\mu \ket{j}$ is the transition dipole moment between states $i$ and $j$ and $F$ is  the field amplitude. 

The relaxation and dephasing processes are introduced in an analogue
way as in Ref.~\cite{PRL1}.  It consists in a 
Liouville equation in Lindblad form to insure complete positivity of
the density matrix describing the quantum system. 
The dynamics of the system is given by:
$\frac{\partial \rho}{\partial t}=\mathcal{L}(t)\rho,$
%\label{eq:LiouvilleEq}
where $\mathcal{L}(t)=\mathcal{L}_H(t)+\so{L}^D$, with
$\hbar\mathcal{L}_H = -i(\openone \otimes H(t) - \bar{H}(t)\otimes
\openone)$ is the Hamiltonian conservative part, with $\bar{H}$ the complex conjugate of $H$, and
 $L^D=L^D_{\text{pop}}+L^D_{\text{pure}}$ is the generator of dissipative  dynamics.
\begin{align}
L^D_{\text{pop}}&=\int dk \Gamma(k) \Big\{ A(k,g)\otimes A(k,g)  \nonumber \\
& - \frac{1}{2}\left[1\otimes A^{\dagger}(k, g)A(k,g) +
  A^{\dagger}(k,g)A(k,g)\otimes 1\right]\Big\}   \nonumber \\
&+\Gamma_e \Big\{ A(e,g)\otimes A(e,g)  \nonumber \\
& - \frac{1}{2}\left[1\otimes A^{\dagger}(e, g)A(e,g) +
  A^{\dagger}(e,g)A(e,g)\otimes 1\right]\Big\},
\label{eq:dissipation_k}
\end{align}
describes the population relaxation from the $\ket{k}$ manifold and from the $\ket{e}$ state to the ground state.
\begin{eqnarray}
&L^D_{\text{pure}} =-\gamma_{eg}\big[\ket{e}\bra{e}\otimes\ket{g}\bra{g}+\ket{g}\bra{g}\otimes\ket{e}\bra{e}\big] \nonumber\\
&-\int dk\gamma_{kg}\big[\ket{k}\bra{k}\otimes\ket{g}\bra{g}+\ket{g}\bra{g}\otimes\ket{k}\bra{k}\big] \nonumber \\
&-\int
dk\gamma_{ke}\big[\ket{k}\bra{k}\otimes\ket{e}\bra{e}+\ket{e}\bra{e}\otimes\ket{k}\bra{k}\big],
\label{eq:pureD}
\end{eqnarray}
describes pure dephasing, that is the dynamics of
the non diagonal matrix elements of $\rho$.
$A(i,j)=\ket{j}\bra{i}$ are the
jump operators and $\Gamma(k)$ is the population relaxation rate from
state $\ket{k}$ to $\ket{g}$ as is $\Gamma_e$ for the $\ket{e}$ population. $\gamma_{ij}$ is the pure dephasing rate for the $ij$ coherence. 
As in Ref.~\cite{PRL1}, we have used the correspondance: $\ket{l}\bra{m} \leftrightarrow \ket{l}\otimes \ket{m}
\equiv\kket{lm}$~\cite{Havel2003}. We use the rotating-wave
approximation (RWA) on
$L=e^{i\underline{\Omega}_Lt}\mathcal{L}(t)e^{-i\underline{\Omega}_Lt}$
and remove non-resonant terms such that $L$ can be considered time
independent. $\underline{\Omega}_L$ is a diagonal matrix whose matrix
elements are equal to  $\pm \omega_L$ for excited(ground)-ground(excited) coherences, and zero elsewhere. 

There are a number of phenomena that occur at strong fields which are not described by the present model (ATI, continuum-continuum transitions). This is in line with previous literature where these processes are also neglected from the model. Within these assumptions, the only restriction of the model is that the Rabi frequency should be much smaller than the laser frequency, which is around the two-level system (TLS) transition as we consider near resonant processes. This is once more an approximation done throughout the literature. This is not restrictive since all the high-field effects described by the model appear at field intensities that are a few percent of the intensities at which the RWA breaks down for a TLS in the visible and a linewidth of around 0.1 eV. In this context, it is the comparison between the laser intensity or more precisely the Rabi frequency and  relaxation rates that determines the limit between the linear and nonlinear regime. 

\section{Feshbach partitioning and effective Liouvillian}

The continuum states population, $\int dk \rho_{kk}$, where
$\rho$ is the full steady state density matrix can be obtained by
finding the kernel of the time-independent operator
$(\underline{\Omega}_L-L)$, that is
%\begin{equation}
$(\underline{\Omega}_L-L)\rho=0.$
%\label{steady-state}
%\end{equation}
To solve this equation, we 
split $L$  in two terms,   $L = L_0 + \mathcal{V}$ where $L_0$ is
diagonal and $\mathcal{V}$  is purely non diagonal, and proceed to Feshbach partitioning \cite{Feshbach1962}. For that, we introduce the projectors $P=\ket{g}\bra{g}+\ket{e}\bra{e}$ and $Q=\int dk \ket{k}\bra{k}$, with $P+Q=1$. The corresponding projectors for the discrete and continuum parts in Liouville space are given by:
\begin{equation}
\mathcal{P}=P\otimes P;\quad%
\mathcal{Q}=P\otimes Q + Q\otimes P + Q\otimes Q.
\label{eq:partitions}
\end{equation}
This allows us to rewrite the kernel equation as
$(\underline{\Omega}_L-L)(\mathcal{P}+\mathcal{Q})\rho=0$. By
projecting on both $\mathcal{P}$ and $\mathcal{Q}$ and after some
algebra (see Appendix A), we obtain
\begin{equation}
\mathcal{P}( \underline{\Omega}_L  - L_{\text{eff}}) \mathcal{P} \rho=0,
\label{rho-discrete}
\end{equation}
with an effective Liouville operator,
\begin{equation}
L_{\text{eff}} = \mathcal{P} L \mathcal{P}+\mathcal{P}\mathcal{V}\mathcal{Q}\mathcal{G}_{\mathcal{Q}}\mathcal{Q}\mathcal{V}\mathcal{P},
\label{Leff}
\end{equation}
and $\mathcal{G}_{\mathcal{Q}}=
(\mathcal{Q}(\underline{\Omega}_L-L)\mathcal{Q})^{-1}$. $\mathcal{G}_{\mathcal{Q}}$,
which is related to the resolvent of $\mathcal{Q}L\mathcal{Q}$,
is not straightforward to calculate unless
$\mathcal{Q}L\mathcal{Q}$ is diagonal. To achieve the calculation of
$\mathcal{G}_{\mathcal{Q}}$, we proceed to a sub-partition of the
$\mathcal{Q}$ subspace until the projected Liouvillian be diagonal (see Appendix A).
$\mathcal{P} \underline{\Omega}_L\mathcal{P}    - L_{\text{eff}}$ acts on the
$\mathcal{P}$ space only, but its kernel  is equal to the projection $P\rho$ of the exact stationary density matrix, on the discrete subspace. 
Once the  density matrix $\mathcal{P}\rho$ on the
discrete space has been obtained,  the  density matrix
$\mathcal{Q}\rho$ in the continuum subspace can be computed
 through the following equation:
\begin{equation}
\begin{split}
\mathcal{Q}\rho&=\mathcal{Q}\mathcal{G}_{\mathcal{Q}}\mathcal{Q} \mathcal{V}\mathcal{P}\rho.
\end{split}
\label{eq:continuum-rho}
\end{equation}
$L_{\text{eff}}$ can be thought as a
$4\times4$ matrix when $\mathcal{P}\mathcal\rho$ is considered as a column vector
with 4 elements. To obtain an explicit expression
for $L_{\text{eff}}$, the usual wide-band approximation is employed. It is
also in this same approximation that an explicit expression was obtained in
the original Fano problem~\cite{Fano1961}.
It
consists in assuming that the parameters of the model do not
depend upon $k$. From now on, we consider this approximation and
define $\Gamma_c\equiv \Gamma(k)$, $\mu_c\equiv\mu_k$ as constants.

After tedious but straightforward calculations (see Appendix A), the effective Liouvillian $L_{\text{eff}}$ defined
in Eq.~\eqref{rho-discrete} can be written in a surprisingly simple form:
\begin{equation}
L_{\text{eff}}=-i(\openone \otimes H_{\text{eff}} - \bar{H}_{\text{eff}}\otimes \openone)+\so{L}_{\text{eff}}^D,
\label{eq:Leff-explicit}
\end{equation}
where $H_{\text{eff}}$ is an effective Hamiltonian and $L_{\text{eff}}^{D}$ is the dissipative part of the effective Liouvillian. We show explicitly in Appendix B that $L_{\text{eff}}$ has a Lindblad form. It can thus be considered as the generator of complete positive evolution. This ensure that its kernel $P\rho$ is indeed a physical state. Here, we prefer a different presentation of the operator that makes it more amenable for comparison to the effective Hamiltonian that is calculated in the scattering problem where dissipation is ignored. The effective Hamiltonian can be written as:
\begin{align}
H_{\text{eff}}&= PH_0P+H_{\text{field}}, \text{with } \\
H_{\text{field}}&=\frac{F}{2}\big(\mu_e-in\pi
V\mu_c\big)\big(\ket{g}\bra{e}+\ket{e}\bra{g}\big).\nonumber
\end{align}
%\begin{align*}
%$PH_0P =\epsilon_0\ket{0}\bra{0}+\epsilon_e\ket{e}\bra{e}$
%\end{align*}
%
It contains an Hermitian and an anti-Hermitian part and corresponds exactly to the effective Hamiltonian of typical resonance problems in Hilbert space \cite{Bertlmann2006,moiseyev2011}. The non-Hamiltonian part of the effective Liouvillian is:
\begin{equation}
\begin{split}
L_{\text{eff}}^D&=L^D_{e}+L^D_{c}+L^D_{\text{pure}}, \text{ with:}\\
\hbar L^D_{e}&= (2n\pi V^2 + \hbar\Gamma_e )\bigg[ A(e,g)\otimes A^{\dagger}(e,g) \\ 
&- \frac{1}{2}\bigg(A^{\dagger}(e,g)A(e,g)\otimes 1 
+ 1 \otimes A^{\dagger}(e,g)A(e,g)\bigg) \bigg] \\
L^D_{\text{pure}} &=-(\gamma_{eg}+n\pi \mu_c^2)\big[\ket{e}\bra{e}\otimes\ket{g}\bra{g}+\ket{g}\bra{g}\otimes\ket{e}\bra{e}\big] \\
\hbar L^D_{c}&= 2n\pi V\mu_c (\kket{gg}\bbra{eg}+\kket{gg}\bbra{ge})
\end{split}
\end{equation}
$L^D_{e}$ is a dissipation superoperator in Lindblad form that describes the
population decay with rate $2n\pi V^2/\hbar+\Gamma_e$ due to the coupling to the continuum and the natural decay rate $\Gamma_e$, and $L^D_{\text{pure}}$ is a
pure dephasing superoperator. $L^D_{c}$ is an additional part of
the relaxation superoperator which cannot be put into a Lindblad form. We stress that this presentation allows a comparison to the Hilbert space solution but that the full $L_{\text{eff}}$ operator of Eq. \eqref{eq:Leff-explicit} can be put in Lindblad form (see Appendix B). 
%Contrary to our previous study~\cite{PRL1}, where pure dephasing
%of the continuum (see Eq.\eqref{eq:pureD}) was mandatory to describe
%fluorescence emission, here, it does not contribute
%to $L_{\text{eff}}$ and therefore will not play any role in state
%populations.

Finally, solving Eqs.~\eqref{rho-discrete} amounts to finding the kernel
of a $4\times4$ matrix, which can be explicitely done with the help of
a symbolic calculation software (see Appendix C). Then, applying Eq.~\eqref{eq:continuum-rho} along
with the normalization condition $\rho_{gg}+\rho_{ee} + \int\ud k\rho_{kk}=1$
gives us all of the populations and coherences. 

The results will be given in terms of dimensionless quantities and
$\hbar\gamma = n\pi V^2$ is taken as the unit of energy.
In addition to the
original dimensionless Fano parameters
$\epsilon=(\omega_L-\omega_e)/\gamma$ and $q=\mu_e/n\pi V \mu_c$, we introduce the new 
parameter $\Omega=\frac{\mu_eF}{2q\hbar \gamma}=\mu_cF/2V$, which corresponds to a dimensionless
Rabi frequency. Also, all relaxation rates will be given in units of
$\gamma$, this amounts to perform the following replacement:  $\Gamma_c\rightarrow \Gamma_c/\gamma$, $\Gamma_e\rightarrow \Gamma_e/\gamma$, $\gamma_{eg}\rightarrow \gamma_{eg}/\gamma$. 

\section{Generalized Fano profile with dissipation}

The main result of the paper is that the population of the excited
state $n_c = \int dk \rho_{kk}$ can  always be brought back to a
Fano  profile $f$,  plus a Lorentzian term:
\begin{equation}
n_c(\epsilon_{\text{eff}};q_{\text{eff}})=C\bigg[f(\epsilon_{\text{eff}};q_{\text{eff}})+\frac{D}{\epsilon_{\text{eff}}^2+1}\bigg],
\label{eq:Effective-equation}
\end{equation}
where the dependence upon $\omega_L$ is solely contained in
$\epsilon_{\text{eff}} =
\frac{\omega_L-\omega_{\text{eff}}}{\gamma_{\text{eff}}}$. $\omega_{\text{eff}}$,
$\gamma_{\text{eff}}$ and $q_{\text{eff}}$ are effective Fano
parameters that depend on all the parameters of the model but
$\omega_L$. $C$ is a measure of the total  population
and $D$ indicates the relative weight of the Lorentzian term in
comparison to the Fano profile. The transformation into the above form involves a rescaling of the parameters achieved in the accompanying software (see Appendix C). 

Simple explicit expressions for the effective Fano parameters can be
given when the
relaxation and the  dephasing rates concerning the $\ket{e}$ state can be
neglected, that is when  $\Gamma_e=0$ and $\gamma_{eg}=0$ (see Eq.~\eqref{eq:Fanoparameter}). This is often a very good approximation in the context of semiconductor
quantum dots or in hybrids consisting of an organic molecule adsorbed on metallic or semiconductor
nanoparticles~\cite{QD-lifetime-1,QD-lifetime-2,turro1991,galoppini2003,Petersson2000} ($1/\Gamma_e \approx \text{nanoseconds}$, $\hbar/n\pi V^2 \approx 10 \text{ femtoseconds}$). 
In that case, the only relaxation process
consists in the continuum states population decay to the ground
state and the profile is given by a pure Fano profile and the
Lorentzian term is absent, $D=0$ and $C=\frac{2\Omega^2}{2\Omega^2+\Gamma_c}$.
We have obtained:
\begin{widetext}
\begin{align}
\frac{\gamma_{\text{eff}}}{\gamma}&=\frac{\Gamma_c\left[1+(q^2+1)\Omega^2(\Omega^2[(2\Omega^2+4)/\Gamma_c+1]+2/\Gamma_c+2)\right]^{1/2}}{2\Omega^2+\Gamma_c}
\nonumber\\
\frac{\omega_{\text{eff}}}{\gamma} &=\frac{\omega_e}{\gamma} +
q\Omega^2\bigg(1-\frac{2}{2\Omega^2+\Gamma_c}\bigg); \quad
\frac{q_{\text{eff}}}{q}= \frac{\Gamma_c}{2\Omega^2+\Gamma_c}\frac{1}{\gamma_{\text{eff}}}.
\label{eq:Fanoparameter}
\end{align}
\end{widetext}

\begin{figure}	
\includegraphics[width=0.45\textwidth]{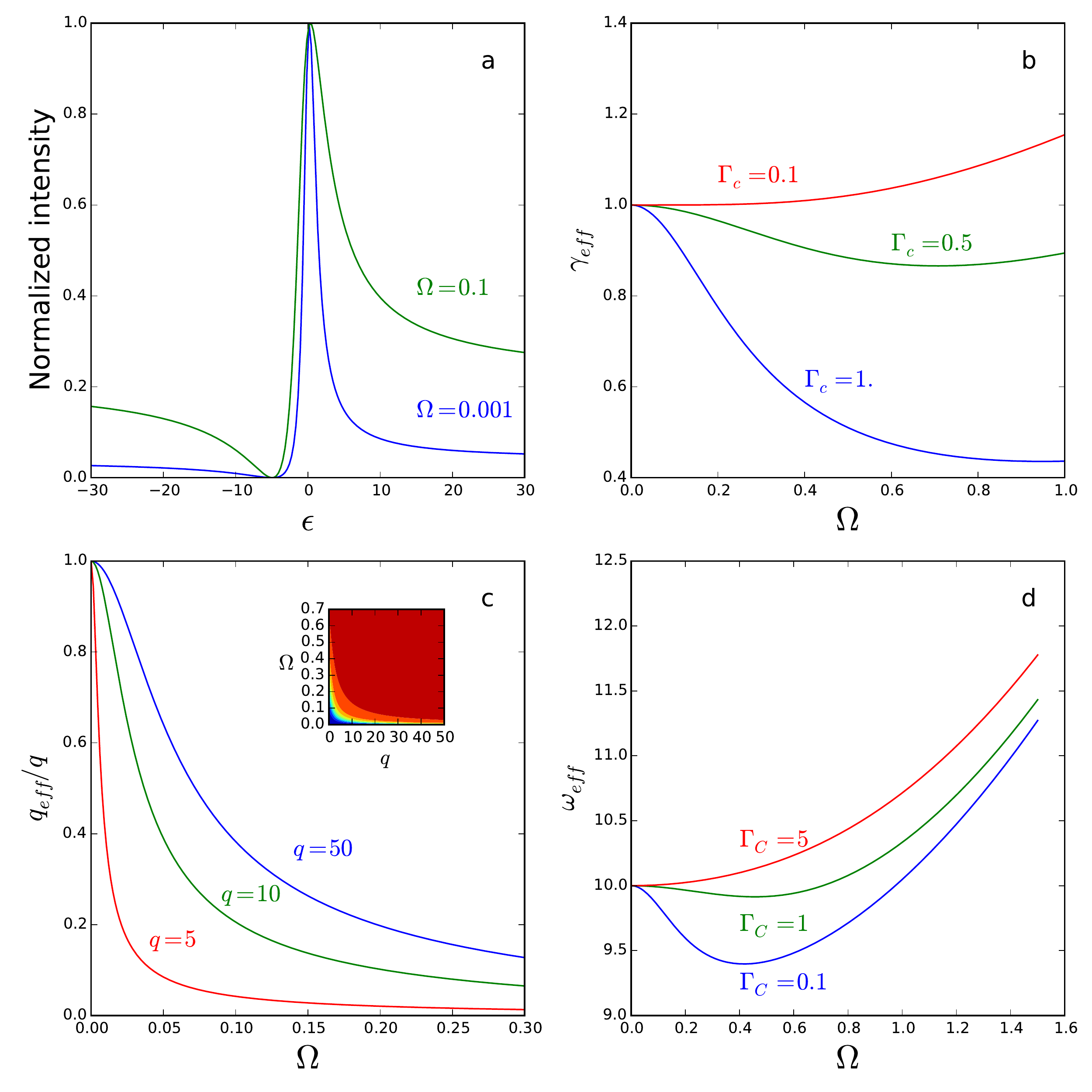}
\caption{ Effect of the field on the Fano profiles and parameters: (a) Fano profiles for $q=5$ for $\Omega=0.001$ and $\Omega=0.1$ (b) $\gamma_{\text{eff}}$ for different values of $\Gamma_c$ (c) $q_{\text{eff}}/q$ for different values of $q$. Inset show $n_c$ as a function of $q$ and $\Omega$. Upper end (red) corresponds to $n_c=1$ and lower end (blue) corresponds to $n_c=0$ (d)  $\omega_{\text{eff}}$ for different values of $\Gamma_c$ and $\omega_e=10$}
\label{fig:4panels}
\end{figure}

We now discuss each one of the parameters as a function of the Rabi frequency and illustrate them in Figure \ref{fig:4panels}. We note that these parameters have a non-linear dependence on the Rabbi frequency $\Omega$, or in other words, a non-linear dependence on the strength of the field.
The prefactor $C$ which is proportional to the intensity of the field for weak fields (linear regime) saturates to $C=1$ when $2\Omega^2 \gg \Gamma_c$. 
In Fig.~\ref{fig:4panels}a, we show the normalized Fano profiles at two intensities of the field ($\Omega=0.001$ and $\Omega=0.1$ for $q=5$). As the field intensity increases, we see changes in all of the Fano parameters. The effective width $\gamma_{\text{eff}}$ increases or decreases (power narrowing or power broadening) depending on the value of the relaxation (see Fig. \ref{fig:4panels}b and Eq.~\eqref{eq:Fanoparameter}). The effective asymmetry parameter $q_{\text{eff}}$ decreases monotonically with $q$ (see Figure \ref{fig:4panels}c and Eq.~\eqref{eq:Fanoparameter}). As shown in the inset, the population of the continuum, even for modest values of $\Omega	$ is significant, underlying the importance of a theory which can handle non-negligible population in the continuum set of states, contrary to the approximations in Ref.~\cite{Zhang2011}. The decrease of $q_{\text{eff}}$ can be thought as a consequence of the  saturation of the discrete excited state population.
The energy shift $\hbar(\omega_{\text{eff}}-\omega_e)$ of the discrete state $\ket{e}$, that can be seen as an AC Stark shift, has an interesting behavior. For $\Gamma_c<2$, the shift is negative if $0<\Omega<\sqrt{(2-\Gamma_c)/2}$ and it is positive if $\Omega>\sqrt{(2-\Gamma_c)/2}$. Therefore, $\Gamma_c<2$, $\Omega=\sqrt{(2-\Gamma_c)/2}$ is a null point.
On the contrary, for large relaxation rates such that $\Gamma_c>2$, the shift will be positive for all values of the field (see Fig. \ref{fig:4panels}d). 

Inclusion of population relaxation  ($\Gamma_e \neq
0$) or  pure dephasing processes ($\gamma_{eg} \neq 0$) 
results into heavy expressions of the states populations that we
provide in the accompanying software (see Appendix C).
The main qualitative features remain unchanged
except for the appearance of a Lorentzian term.

% Its relative amplitude $D$ can be written explicitly when $\Gamma_e \neq 0$ and $\gamma_{eg}=0$:
%\begin{equation}
%D=\frac{\Gamma_e(\Omega^2+\Gamma_e+1)[(q^2+\Gamma_e+1)\Omega^2+\Gamma_e(1+\Gamma_e)]}{(1+\Gamma_e)^2}.
%\end{equation}

Until now, we have focused on the stationary population of the
continuum set of states $\int\ud k \rho_{kk}$. 
%This quantity reflects
%the absorption rate of  a monochromatic  light. To be more precise, the
%absorption rate is given by $\Gamma_c \int\ud k \rho_{kk} +
%\Gamma_e\rho_{ee}$.
Another quantity that can be measured is the photocurrent, that is the
total flow of electrons in the continuum. Assuming that all the
electrons emitted in the continuum $\ket{k}$ states are collected by an electrode, in the limit of low bias
voltage, the current intensity $I$ can be obtained from the stationary
populations as $I=\lim_{\Gamma_c \rightarrow 0}\abs{e}\frac{\Gamma_c \int dk
  \rho_{kk}}{\rho_{gg}}$~\cite{Davis1997}, where $e$ is the electron
charge. It turns out that $I$ can also be brought into the form of a Fano factor and a
Lorentzian factor as in
equation~\eqref{eq:Effective-equation}. Explicit expressions for the
Fano and Lorentzian parameters can be given with both the population relaxation and
dephasing included:

\begin{widetext}
\begin{align}
%\begin{split}
\gamma_{\text{eff}}^{\text{tr}} &=
(1+\Gamma_e)^{-1}\left[\Omega^4\Gamma_e(q^2+\Gamma_e+1) +\Omega^2(1+\Gamma_e)(q^2+2\Gamma_e+1)(\Gamma_e+\gamma_{eg}+1)
+ (1+\Gamma_e)^2(\Gamma_e+\gamma_{eg}+1)^2\right]^{1/2} \\
D^{\text{tr}}&=\frac{1}{(\gamma_{\text{eff}}^{\text{tr}})^2}(1+\Gamma_e)^{-2}\left[\Omega^4\Gamma_e(q^2+\Gamma_e+1) +\Omega^2(1+\Gamma_e)(q^2+2\Gamma_e+1)(\Gamma_e+\gamma_{eg})\right]
\nonumber \\
& +\frac{1}{(\gamma_{\text{eff}}^{\text{tr}})^2}(1+\Gamma_e)^{-1}\left[\Gamma_e^3+\Gamma_e^2+\gamma_{eg}(2\Gamma_e^2+\Gamma_e
  \gamma_{eg}+q^2+2\Gamma_e+\gamma_{eg}+1)\right],
%\end{split}
\label{eq:Dtr}
\end{align}
\end{widetext}
\begin{equation}
\frac{\omega_{\text{eff}}^{\text{tr}}}{\gamma} =
\frac{\omega_e}{\gamma}+\frac{q\Omega^2}{1+\Gamma_e}; \quad
q_{\text{eff}}^{\text{tr}}=\frac{1}{\gamma_{\text{eff}}^{\text{tr}}
}; \quad
C^{\text{tr}}=2\Omega^2.
\end{equation}

These expressions give an exact description of the scattering
problem, as formulated by Fano in its original work~\cite{Fano1961}, but extended to large
field intensities and dissipation processes.

\begin{figure}	\includegraphics[width=0.35\textwidth]{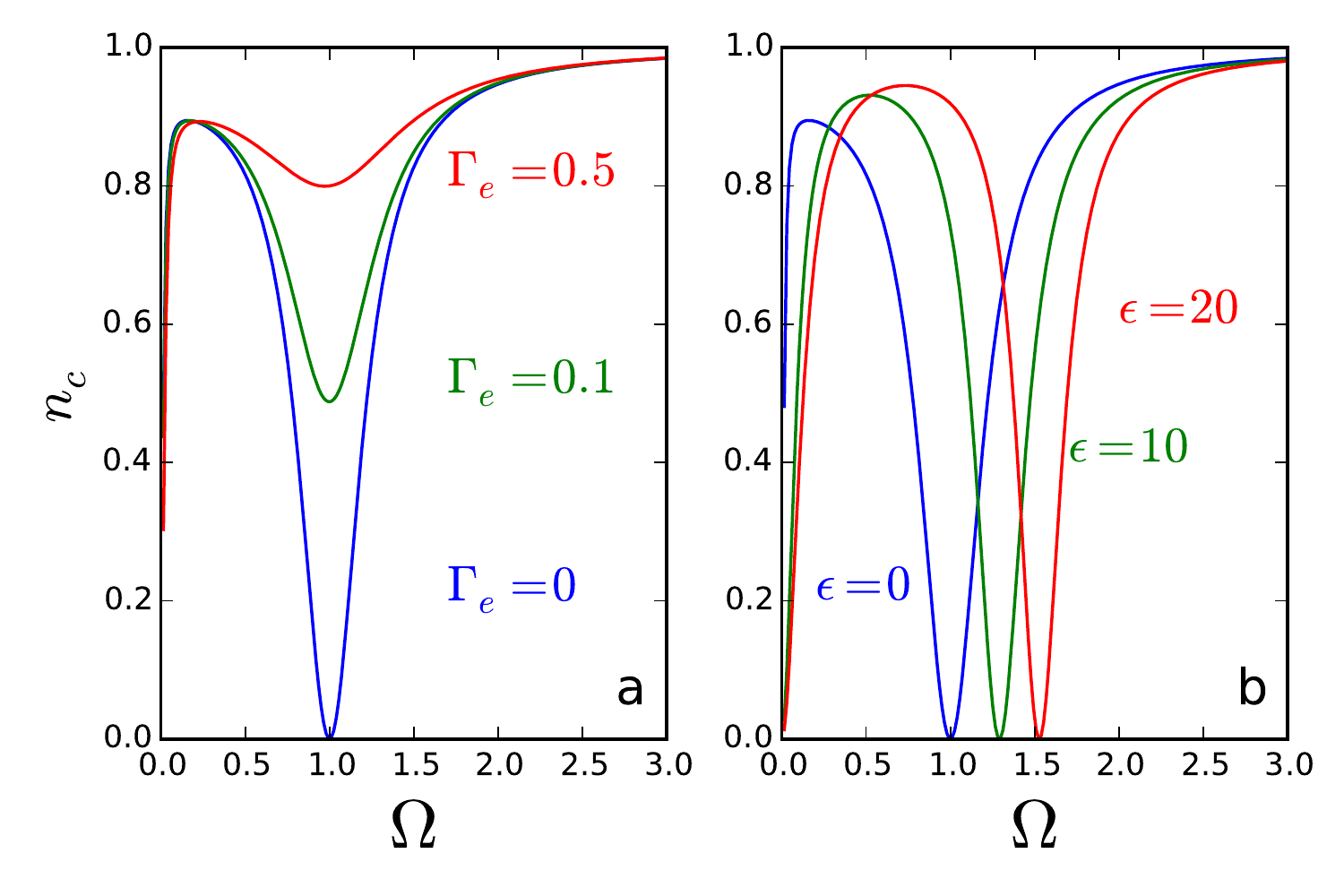}
	\caption{Population in the continuum showing electromagnetically induced transparency (EIT) for a) $\epsilon=0$, $q=15$ with different values of $\Gamma_e$ and b) for $q=15$, $\Gamma_e=0$ different values of $\epsilon$}  
\label{fig:EIT}
\end{figure}

An interesting consequence of the nonlinear Fano
effect, is electromagnetically-induced transparency (EIT)~\cite{Boller1991}. Indeed, in the absence of
discrete state relaxation and dephasing  ($\Gamma_e=\gamma_{eg}=0$), then $D^{\text{tr}}=0$ in Eq.~\eqref{eq:Dtr} and therefore the continuum population goes through zero when $\epsilon_{\text{eff}} =-q_{\text{eff}}$ (see Eq.~\eqref{eq:Effective-equation}). 
The phenomenon of EIT has been characterized before \cite{Buth2007,Glover2010}. In the standard scheme, EIT is obtained with two laser frequencies, where one acts as the control radiation that creates the transparency window while the second one acts as a probe. In our case, which is non-standard, the same frequency acts as a control and probe radiation, and the transparency window arises from two pathways whose destructive interference point is tunable via its intensity. 
It can be shown that the condition $\epsilon_{\text{eff}}=-q_{\text{eff}}$ is equivalent to
$\Omega^2=1+\frac{\epsilon}{q}$ either for the light absorption or the photocurrent
intensity. This phenomenon is shown in Fig.~\ref{fig:EIT}.a for the
case $\epsilon=0$, $q=15$ for different values of $\Gamma_e$, and $q=15$, and Fig.~\ref{fig:EIT}.b $\Gamma_e=0$ for different values of $\epsilon$.
 This phenomenon provides an interesting tool for devices as well as a
 means for determining the system parameters. For example, irradiating
 at the discrete level resonance ($\epsilon=0$) in weak field and
 increasing the intensity until the induced transparency is found
 determines the ratio of transition dipole moment $\mu_c$ to the
 coupling $V$ such that $\mu_c/2V=1/F$. In  the presence of pure
 dephasing or of relaxation from the discrete state, the zero becomes
 a minimum but its position  does not change appreciably (see Figure \ref{fig:EIT} (left)). 
 One should take care not to confuse the present effect  with the trivial zero of the standard Fano profile at $\epsilon=-q$ (see Eq. \eqref{eq:Fano-classic}). In the case presented here, the effect is non-linear as the condition is $\epsilon_{\text{eff}}=-q_{\text{eff}}$ and both of these parameters depend non-linearly on the field. The condition for the minimum is thus obtained by adjusting the intensity of the field. 

\section{Conclusion}

We have obtained an explicit formula solving the original Fano problem for arbitrary relaxation processes and large radiative couplings. The inclusion of more than a single excited discrete state is straightforward as long as the couplings do not exceed the energy gap between excited states. Furthermore, our approach serves as a stepping stone for descriptions going beyond the wide-band approximation, already discussed in the scattering framework. Finally, there is a wide class of systems where the present  model is directly applicable opening new horizons in the analysis of Fano profiles under intense fields, as well as in applications and devices that exploit processes such as population inversion and electromagnetically induced transparency.

\textbf{Acknowledgements} D.F.S. acknowledges the Research in Paris program for a fellowship and Victoria Cantoral Farfan and Sadek Salem Al Harbat (UPMC) for fruitful discussions. This work was financially supported by project NSF-ANR (ANR-11-NS04-0001 FRAMOLSENT program, NSFCHE-112489). This work was performed using HPC resources from GENCI- CINES/IDRIS (Grant 2015- x2015082131, 2014- x2014082131) and the CCRE-DSI of Universit\'e P. M. Curie.

\bibliography{Fano}

\section{Appendix A: Steady-state density matrix}

We outline the procedure used to obtain the steady-state density matrix by finding the kernel of $\underline{\Omega}_L-L$. This process can be separated into two different equations, one for the discrete subspace and one for the continuum subspace. The density matrix in the discrete subspace $\mathcal{P}\rho$ is the solution to:
\begin{equation}
[\Omega_L-(\mathcal{P} L \mathcal{P}+\mathcal{P}\mathcal{V}\mathcal{Q}\mathcal{G}_{\mathcal{Q}}\mathcal{Q}\mathcal{V}\mathcal{P})]\rho=0
\end{equation}
The density matrix in the continuum subspace $\mathcal{Q}\rho$ can then be calculated:
\begin{equation}
\mathcal{Q}\rho=\mathcal{Q}\mathcal{G}_{\mathcal{Q}}\mathcal{Q} \mathcal{V}\mathcal{P}\rho.
\end{equation}
The derivation of these equations is presented below. Solving them requires the calculaton of the kernel of a $4\times 4$ matrix, which we will do with a symbolic calculator and knowledge of the resolvent $\mathcal{Q}\mathcal{G}_{\mathcal{Q}}\mathcal{Q}=(z+\mathcal{Q}(\Omega_L-L)\mathcal{Q})^{-1}|_{z=0}$. In what follows we will omit $z$ altogether since all of the expressions for the density matrix require evaluation of the resolvent for $z=0$. There are problems arising from this inversion because the operator in parenthesis is infinite dimensional with a continuous spectrum. However, we can partition this equation successively until we arrive at a partition where the Liouvillian is diagonal. The resolvent of a diagonal matrix is trivial and this will be the cornerstone of the rest of the calculation which will use exact resummations of perturbative expansions given by the Lippman-Schwinger equation. \newline

In the case of a Fano-type model, we use three partition spaces which are schematically represented in Figure \ref{fig:Partition}. 
\begin{itemize}
\item $\so{P}=\kket{gg}\bbra{gg}+\kket{ge}\bbra{ge}+\kket{eg}\bbra{eg}+\kket{ee}\bbra{ee}$
\item $\so{P}_2=\int dk \kket{kg}\bbra{kg}+\kket{gk}\bbra{gk}$
\item $\so{P}_3=\int dk \kket{ke}\bbra{ke}+\kket{ek}\bbra{ek}$
\item $\so{Q}_3=\int dk \int dk' \kket{kk'}\bbra{kk'}$
\end{itemize}

\def\tikz@delimiter#1#2#3#4#5#6#7#8{%
  \bgroup
    \pgfextra{\let\tikz@save@last@fig@name=\tikz@last@fig@name}%
    node[outer sep=0pt,inner sep=0pt,draw=none,fill=none,anchor=#1,at=(\tikz@last@fig@name.#2),#3]
    {%
      {\nullfont\pgf@process{\pgfpointdiff{\pgfpointanchor{\tikz@last@fig@name}{#4}}{\pgfpointanchor{\tikz@last@fig@name}{#5}}}}%
      \delimitershortfall\z@% as suggested by morbusg (maximum space not covered by a delimiter = 0)
      \resizebox*{!}{#8}{$\left#6\vcenter{\hrule height .5#8 depth .5#8 width0pt}\right#7$}%
    }
    \pgfextra{\global\let\tikz@last@fig@name=\tikz@save@last@fig@name}%
  \egroup%
}

\tikzset{
  withparens/.style = {draw, outer sep=0pt,
    left delimiter=(, right delimiter=),
    above delimiter=(, below delimiter=),
    align=center},
  withbraces/.style = {draw, outer sep=0pt,
    left delimiter=\{, right delimiter=\},
    above delimiter=\{, below delimiter=\},
    align=center}
}

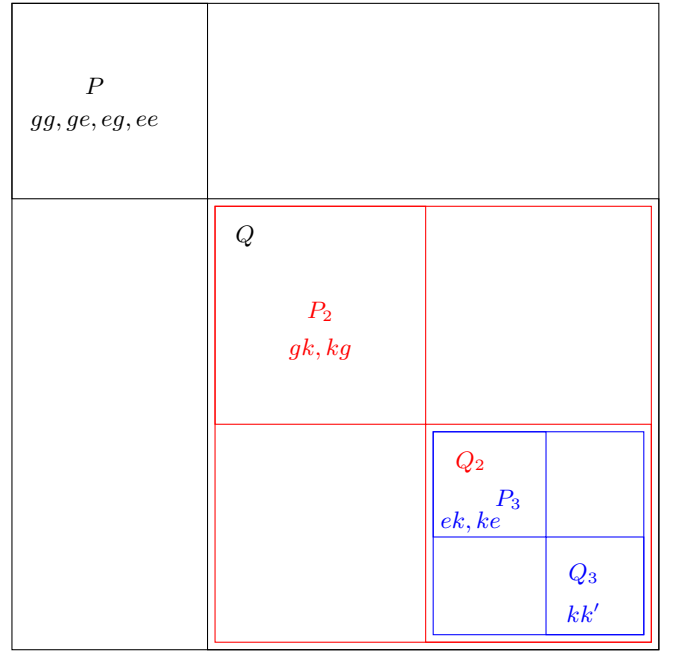
\begin{figure}[ht]
\centering
\begin{tikzpicture}
%First partition
\draw (0cm,0cm) rectangle (8.6cm,8.6cm);
\draw (2.6cm,0cm) rectangle (8.6cm,6cm);
\draw (0cm,6cm) rectangle (2.6cm,8.6cm);
\node at (1.1,7.5) {$\so{P}$};
\node at (1.1,7.0) {$gg,ge,eg,ee$};
\node at (3.1,5.5) {$\so{Q}$};
%Second partition
\draw [red] (2.7cm,0.1cm) rectangle (8.5cm,5.9cm);
\draw [red] (5.5cm,5.9cm) rectangle (2.7cm,3cm);
\draw [red] (5.5cm,0.1cm) rectangle (8.5cm,3cm);
\node at (4.1,4.5) {\color{red} $\so{P}_2$};
\node at (4.1,4.0) {\color{red} $gk,kg$};
\node at (6.1,2.5) {\color{red} $\so{Q}_2$};
%Third partition
\draw [blue] (5.6cm,0.2cm) rectangle (8.4cm,2.9cm);
\draw [blue] (5.6cm,1.5cm) rectangle (7.1cm,2.9cm);
\draw [blue] (7.1cm,0.2cm) rectangle (8.4cm,1.5cm);
\node at (6.6,2) {\color{blue} $\so{P}_3$};
\node at (6.1,1.7) {\color{blue} $ek,ke$};
\node at (7.6,1) {\color{blue} $\so{Q}_3$};
\node at (7.6,0.5) {\color{blue} $kk'$};
\end{tikzpicture}
\caption{\label{fig:Partition} {\it Partition of the Liouvillian superoperator}}
\end{figure}

The first partition divides the continuum from the discrete states and the remaining partitions divide the continuum subspace such that $\so{P}_2$, $\so{P}_3$ and $\so{Q}_3$ are diagonal. 
A Lippman-Schwinger recursion equation applied to the last partition $\so{Q}_3$ and $\so{P}_3$ gives $\so{Q}_2\mathcal{G}_{\mathcal{Q}_2}\so{Q}_2$, a Lippman-Schwinger equation applied to the second partition gives $\so{Q}\mathcal{G}_{\mathcal{Q}}\so{Q}$ and a third and final Lippman-Schwinger equation yields the resolvent in the entire Liouville space $\mathcal{G}$. 

Taking the upper partition as an example, we show the expressions to calculate the resolvent. We assume that $\mathcal{QG}_0\mathcal{Q}=(\mathcal{Q}(\underline{\Omega}_L-L)\so{Q})^{-1}$ and $\mathcal{PG}_0\mathcal{P}=(\mathcal{P}(\underline{\Omega}_L-L)\mathcal{P})^{-1}$ are known from the previous recursion step. We also write the non-diagonal part as $\mathcal{P}L\mathcal{Q}+\mathcal{Q}L\mathcal{P}=\mathcal{V}$. We decompose the Lippman-Schwinger equation $\mathcal{G}=\mathcal{G}_0+\mathcal{G}_0\mathcal{VG}$ into its partitions. We insert $1=\mathcal{P}+\mathcal{Q}$ on both sides of $\mathcal{V}$ and project the entire expressions onto $\mathcal{P}$ and $\mathcal{Q}$. After some rearranging:

\begin{equation}
\begin{split}
&\mathcal{P}\mathcal{G}\mathcal{P}=\mathcal{P}\mathcal{G}_0\mathcal{P}+\mathcal{P}\mathcal{G}_0\mathcal{P}(\mathcal{P}\mathcal{V}\mathcal{Q}\mathcal{G}_0\mathcal{Q}\mathcal{V}\mathcal{P})\mathcal{P}\mathcal{G}\mathcal{P} \\
&\mathcal{Q}\mathcal{G}\mathcal{P}=\mathcal{Q}\mathcal{G}_0\mathcal{Q}\mathcal{V}\mathcal{P}\mathcal{G}\mathcal{P} \\
&\mathcal{P}\mathcal{G}\mathcal{Q}=\mathcal{P}\mathcal{G}\mathcal{P}\mathcal{P}\mathcal{V}\mathcal{Q}\mathcal{Q}\mathcal{G}_0\mathcal{Q} \\
&\mathcal{Q}\mathcal{G}\mathcal{Q}=\mathcal{Q}\mathcal{G}_0\mathcal{Q}+\mathcal{Q}\mathcal{G}_0\mathcal{Q}\mathcal{V}\mathcal{P}\mathcal{G}\mathcal{P}\mathcal{V}\mathcal{Q}\mathcal{G}_0\mathcal{Q}
\end{split}
\label{eq:solution1}
\end{equation}

We can explain more explicitly, why equations \eqref{eq:solution1} solves the problem: the first equation give $\mathcal{PGP}$ by the inversion (or an infinite resumation) in $\mathcal{P}$ space-we can give the explicit equation, and the other equations  give the others projections  in term of $\mathcal{PGP}$.

In our aim to obtain the steady-state density matrix, we do not need to calculate $\mathcal{PGP}$ but only $\mathcal{P}\mathcal{V}\mathcal{Q}\mathcal{G}_0\mathcal{Q}\mathcal{V}\mathcal{P}$. The steady-state density matrix is obtained by $(\underline{\Omega}_L-L)\rho=0$. We can project onto $\mathcal{P}$ and $\mathcal{Q}$:
\begin{equation}
\begin{split}
\mathcal{P}(\underline{\Omega}_L-L)\mathcal{P}\rho+\mathcal{P}(\underline{\Omega}_L-L)\mathcal{Q}\rho&=0 \\
\mathcal{Q}(\underline{\Omega}_L-L)\mathcal{P}\rho+\mathcal{Q}(\underline{\Omega}_L-L)\mathcal{Q}\rho&=0 \\
\end{split}
\end{equation}
which can be rearranged to get:
\begin{equation}
\begin{split}
&\big[\mathcal{P}(\underline{\Omega}_L-L)\mathcal{P}\\
&-\mathcal{P}\mathcal{V}\mathcal{Q}(\mathcal{Q}(\underline{\Omega}_L-L)\mathcal{Q})^{-1}\mathcal{Q}\mathcal{V}\mathcal{P}\big]\rho=0 
\end{split}
\label{Prho}
\end{equation}
and
\begin{equation}
\mathcal{Q}\rho=\big[(\mathcal{Q}(\underline{\Omega}_L-L)\mathcal{Q})^{-1}\mathcal{Q}\mathcal{V}\mathcal{P}\big]\rho
\end{equation}
We recognize that we can group the terms in Equation \eqref{Prho} into an effective Liouvillian $L_{\text{eff}}$ 
\begin{equation}
L_{\text{eff}} = \mathcal{P} L \mathcal{P}+\mathcal{P}\mathcal{V}\mathcal{Q}\mathcal{G}_{\mathcal{Q}}\mathcal{Q}\mathcal{V}\mathcal{P}
\end{equation}
where $\mathcal{G}_{\mathcal{Q}}=
(\mathcal{Q}(\underline{\Omega}_L-L)\mathcal{Q})^{-1}$, leading to expression (5) and (6) of the main text.
To illustrate the procedure we show the calculation of the resolvent for the subspace $\mathcal{P}_3+\mathcal{Q}_3=\mathcal{Q}_2$. We start with the first line of Eq.~\eqref{eq:solution1} which gives the solution for the resolvent in the $\mathcal{P}_3\mathcal{P}_3$ subspace. We can rewrite it in the form of an infinite series $\mathcal{P}_3\mathcal{G}\mathcal{P}_3=\mathcal{P}_3\mathcal{G}_0\mathcal{P}_3\sum_{n=0}^{\infty}(\mathcal{P}_3\mathcal{V}\mathcal{Q}_3\mathcal{G}_0\mathcal{Q}_3\mathcal{V}\mathcal{P}_3\mathcal{G}_0\mathcal{P}_3)^n$. We denote the argument which is exponentiated inside the sum as $w_3$. 

\begin{equation}
\begin{split}
w_3=&-\int dk V^2g(k,k')g(k,e)\kket{ke}\bbra{ke}\\
&-\int dk V^2g(k',k)g(e,k)\kket{ek}\bbra{ek} \\
&\int dk \int dk' V^2g(k',k)g(e,k)\kket{k'e}\bbra{ek}\\
&\int dk \int dk' V^2g(k,k')g(k,e)\kket{ek'}\bbra{ke}
\end{split}
\label{eq:WG-3}
\end{equation}
where $g(a,b)=[-i((\underline{\Omega}_L)_{ab}-E_a+E_b)+\Gamma_{ab}]^{-1}$, and $\Gamma_{ab}$ is the dissipative term of the $\kket{ab}$ element of the density matrix. We need to take the geometric series of term $w_3$. In the wideband approximation, the only terms which will contribute to the final result give for $\mathcal{P}_3(\mathcal{Q}_2(\underline{\Omega}_L-L)\mathcal{Q}_2)^{-1}\mathcal{P}_3$:
\begin{equation}
\begin{split}
\mathcal{P}_3(\mathcal{Q}_2(\underline{\Omega}_L-L)\mathcal{Q}_2)^{-1}\mathcal{P}_3&=\frac{g(k,e)}{1+n\pi V^2g(k,e)}\kket{ke}\bbra{ke}\\
&+\frac{g(e,k)}{1+n\pi V^2g(e,k)}\kket{ek}\bbra{ek}
\end{split}
\end{equation}
this amounts to a renormalization which introduces an effective dissipation term - and thus linewidth - of $n\pi V^2$ to the $ek$ and $ke$ coherences due to the coupling of the discrete excited state with the continuum set of states. This term appears in the final effective Liouvillian as a Lindblad dissipation from the discrete excited state to the ground state with rate $2n\pi V^2/\hbar$. For a more detailed description of the integrals and their evaluation see Supplemental Information of~\cite{PRL1}. \newline

\section{Appendix B: Lindblad form of the effective Liouvillian}

We now write the effective Liouvillian (Eq. \eqref{eq:Leff-explicit}) in Lindblad form. Lindblad \cite{Lindblad1976} and Gorini, Kowassaowki and Sudarshan \cite{Gorini1976} showed that the most general form of a Markovian process was described by a semigroup. This ensures complete positivity and trace-preserving properties of the dynamical map. The Lindblad-GKS form is:
\begin{equation}
L=-i[H,\rho]+\frac{1}{2}\sum^{N^2-1}_{i,j=1}c_{ij}\big(  [F_i,\rho F_j^*]+[F_i\rho ,F_j^*]  \big)
\label{eq:Lindblad}
\end{equation}
where $F_i$'s form a traceless orthonormal set, $H$ is a self-adjoint operator and $c_{ij}$ is a positive definite matrix. We choose as the basis the Pauli matrices:
\begin{equation}
\begin{split}
\sigma_x&=\frac{1}{\sqrt{2}}\begin{bmatrix}
0 & 1 \\
1 & 0
\end{bmatrix},\: \sigma_y=\frac{1}{\sqrt{2}}\begin{bmatrix}
0 & i \\
-i & 0
\end{bmatrix},\: \sigma_z=\frac{1}{\sqrt{2}}\begin{bmatrix}
1 & 0 \\
0 & -1
\end{bmatrix}
\end{split}
\end{equation}
and express the effective Liouvillian in the new basis.
The result is:
\begin{equation}
\begin{split}
L_{\text{eff}}&=L_0+\mathcal{PVQG_Q}(0)\mathcal{QVP} \\
L_{\text{eff}}-L_0&=-\frac{i}{\hbar}[H',]+L^D \\
H'&=\sqrt{2}n\pi V\mu_c \sigma_y  \\
\end{split}
\end{equation}
where $L_0$ is the original Liouvillian in the $\mathcal{P}$ partition, $H'$ is an (Hermitian) addition to the Hamiltonian and $L^D$ is a superoperator in Lindblad form (Eq. \eqref{eq:Lindblad}). The strictly dissipative part can be completely determined by the coefficients $c_{ij}$:
\begin{equation}
c_{ij}=\begin{bmatrix}
c_3 & ic_3 & c_2 \\
-ic_3 & c_3 & -ic_2\\
c_2 & ic_2 & c_1 
\end{bmatrix}
\end{equation}
with $c_1=\Omega^2+\gamma_{eg}$, $c_2=\Omega$ and $c_3=1+\Gamma_e$.
To check the positivity of the $c_{ij}$ matrix we diagonalize it. The eigenvalues are:
\begin{equation}
\begin{split}
c_{\pm}&=\frac{1}{2}\bigg[(\Omega^2+2\Gamma_e+\gamma_{eg}+2)\\
& \pm \sqrt{(\Omega^2-2\Gamma_e+\gamma_{eg}-2)^2-8 \Omega^2 }\bigg]
\end{split}
\end{equation}
Both eigenvalues are positive proving that the matrix is positive definite and that the operator can in fact be written in Lindblad form. 

\section{Appendix C: Accompanying software}

To make the results of this paper more accessible, we have included all results, equations and plots in an interactive python notebook. The notebook consists of a symbolic part where the expressions that include dissipation from the discrete state and pure decoherence - too intricate to write in an article - can be obtained. The second part consists of numerical simulations where the plots produced in the paper can be redrawn and where any simulation of the solution to the Fano model can be plotted. The software can be downloaded from finoqs.wordpress.com. 

\end{document}